\documentclass{article}

\usepackage{PRIMEarxiv}

\usepackage[utf8]{inputenc} 
\usepackage[T1]{fontenc}    
\usepackage{hyperref}       
\usepackage{url}            
\usepackage{booktabs}       
\usepackage{amsfonts}       
\usepackage{nicefrac}       
\usepackage{microtype}      
\usepackage{fancyhdr}       
\usepackage{graphicx}       
\graphicspath{{media/}}     

\usepackage{graphicx}%
\usepackage{multirow}%
\usepackage{amsmath,amssymb}%
\usepackage{amsthm}%
\usepackage{mathrsfs}%
\usepackage[title]{appendix}%
\usepackage{xcolor}%
\usepackage{textcomp}%
\usepackage{float}
\usepackage{manyfoot}%
\usepackage{algpseudocode}%
\usepackage{algorithm2e}
\floatstyle{ruled}
\newfloat{algorithm}{tbp}{loa}
\floatname{algorithm}{Table}
\usepackage{listings}%

\usepackage{array}
\newcolumntype{P}[1]{>{\centering\arraybackslash}p{#1}}
\newcolumntype{M}[1]{>{\centering\arraybackslash}m{#1}}
\newcolumntype{L}{>{\raggedleft} p{0.225\textwidth}}
\newcolumntype{R}{>{\raggedright}p{0.725\textwidth}}
\usepackage{rotating}
\usepackage{lscape}

\title{Spatially-clustered spatial autoregressive models with application to agricultural market concentration in Europe
}

\author{
  Roy Cerqueti \\
  Department of Social and Economic Sciences  \\
  Sapienza University of Rome, Italy \\
  \& \\
  GRANEM \\
  University of Angers, France \\
  \texttt{roy.cerqueti@uniroma1.it} \\
   \And
  Paolo Maranzano \\
  Department Economics, Management and Statistics (DEMS) \\ 
  University of Milano-Bicocca, Italy \\
  \texttt{paolo.maranzano@unimib.it} \\
  \& \\
  Fondazione Eni Enrico Mattei (FEEM) \\
  Milano, Italy \\
  \AND
  Raffaele Mattera \\
  Department of Social and Economic Sciences  \\
  Sapienza University of Rome, Italy \\
  \texttt{raffaele.mattera@uniroma1.it} \\
}

\begin{document}
\maketitle

\begin{abstract}
In this paper, we present an extension of the spatially-clustered linear regression models, namely, the spatially-clustered spatial autoregression (SCSAR) model, to deal with spatial heterogeneity issues in clustering procedures. In particular, we extend classical spatial econometrics models, such as the spatial autoregressive model, the spatial error model, and the spatially-lagged model, by allowing the regression coefficients to be spatially varying according to a cluster-wise structure. Cluster memberships and regression coefficients are jointly estimated through a penalized maximum likelihood algorithm which encourages neighboring units to belong to the same spatial cluster with shared regression coefficients. Motivated by the increase of observed values of the Gini index for the agricultural production in Europe between 2010 and 2020, the proposed methodology is employed to assess the presence of local spatial spillovers on the market concentration index for the European regions in the last decade. Empirical findings support the hypothesis of fragmentation of the European agricultural market, as the regions can be well represented by a clustering structure partitioning the continent into three-groups, roughly approximated by a division among Western, North Central and Southeastern regions. Also, we detect heterogeneous local effects induced by the selected explanatory variables on the regional market concentration. In particular, we find that variables associated with social, territorial and economic relevance of the agricultural sector seem to act differently throughout the spatial dimension, across the clusters and with respect to the pooled model, and temporal dimension.
\end{abstract}


\section{Introduction} \label{sec1}

In social and economic sciences, understanding the relationships between variables is essential for making informed decisions and developing effective policies. Traditional regression models are commonly used for this purpose, which however often assume a uniform relationship across all the statistical units in the dataset. This assumption may not hold in many real-world problems, where different groups of units may exhibit distinct relationships between variables. This phenomenon is known as the Simpson's paradox in statistical literature \cite{alin2010simpson}. Fail to account for clustering structure may lead to severe bias in the estimates.

Clusterwise regression \cite{desarbo1988maximum}, also known as k-means regression \cite{spath1979algorithmus,hathaway1993switching}, addresses the limitation of standard regression models by acknowledging the potential heterogeneity in the relationships across different clusters. By identifying and modeling these heterogeneous relationships, clusterwise regression models iteratively estimate both the unknown clustering structure in the data and the within-cluster regression coefficients. Thus, this class of methods provides a deeper analysis of the data, allowing researchers to uncover hidden patterns, tailor interventions, and better understand complex systems. 

\cite{spath1979algorithmus} proposed one of the first clusterwise regression algorithms, alternating parameter estimation step and residuals-based clustering, which has been adopted in many other studies (e.g. see, \cite{ari2002clustered,bagirov2017prediction,brusco2003multicriterion,wedel1989consumer}) and extended to fuzzy clustering setting by \cite{hathaway1993switching} and \cite{coppi2006least}, among the others. \cite{desarbo1988maximum} proposed a Gaussian mixture-based clustering regression approach, whose properties have been studied in more detail by \cite{hennig2000identifiablity} and \cite{brusco2008cautionary}, while \cite{sugasawa2021grouped} recently proposed a more general parametric approach for cross-sectional data. \cite{di2023lasso} introduced a variable selection step in a similar framework through LASSO-type shrinkage.

When dealing with georeferenced statistical units, however, the spatial dimension needs to be considered in regression modeling. These units are inherently interconnected with their neighboring areas, as dictated by the first law of geography \cite{tobler1970computer}. Spatial clustering, used to identify clusters of units based on geographic proximity, has to be considered. Incorporating spatial clustering into clusterwise regression analysis allows researchers to account for spatial effects effectively. In the framework of clusterwise regression, the spatial dimension has recently been introduced in \cite{sugasawa2021spatially}, extending the mixture regression approach to incorporate spatial clustering by the use of a \cite{potts1952some}-type spatial penalty term. The idea to incorporate spatial clustering into regression has also been proposed in other studies (e.g., \cite{bille2017two,lee2017cluster}, which however mainly developed two-step procedures. Differently, \cite{sugasawa2021spatially} carried out regression and clustering simultaneously. Therefore, the last is the type of approach considered in this paper.

However, we notice that the use of spatial clustering in the clusterwise regression model enhances the presence of spatial autocorrelation in the within-clusters residuals. This violates the strong assumption of iid error terms in traditional clusterwise regression models and can lead to biased estimates of regression coefficients and inaccurate inference \cite{anselin2013spatial}. Moreover, inaccurate estimation of the regression parameters leads to a biased computation of the residuals, which is the main driver of the clustering updating mechanism in the clusterwise algorithms. To address this issue in standard regression models, researchers have proposed incorporating spatial lags in the dependent variable, error term, and covariates into the regression model \cite{elhorst2014spatial}.

In this paper, we propose a solution to this still open issue in the clusterwise regression literature, that is, an extension of the spatially-clustered linear regression models called the spatially-clustered spatial autoregression models (SCSAR). We extend a classical spatial econometrics model, the spatial autoregressive model (SAR), by allowing the regression coefficients to be spatially varying according to a cluster-wise structure. We also discuss the extension to the spatially clustered setting of other spatial models such as the spatial error model (SEM) and the spatial lagged covariates (SLX) model.  Cluster memberships and regression coefficients are jointly estimated through a penalized maximum likelihood algorithm which encourages neighboring units to belong to the same spatial cluster with shared regression coefficients. 

The proposed methodology is then applied to investigate the structural changes that occurred in the agricultural industry in Europe in the last decade. In particular, available data on the number of active farms and their economic capacity and production (standard output) revel that between 2010 and 2020 the agricultural market suffered from a marked increase in its concentration as the number of small and micro farms decreased and the count of large holdings (i.e., with high volumes of production) increased sensibly. Such a concentration process is testified by increasingly values of the Gini index for the standard output throughout Europe. The SCSAR model is applied to a spatio-temporal dataset aiming at assessing the presence of local spatial spillovers of market concentration in the European regions (NUTS-2 level data) in 2010 and 2020.
Empirical findings support the hypothesis of a marked fragmentation of the European agricultural market, which can be well represented by a clustering structure partitioning the whole are into three-groups, roughly approximated by a division among Western (i.e., France and the Iberian peninsula), North-Central (i.e., Mittel-European and Scandinavian regions) and South-Eastern regions. Also, we detect heterogeneous local effects induced by the selected explanatory variables on the regional market concentration. In particular, we find that variables associated with social, territorial and economic relevance of the agricultural sector seem to act differently throughout the spatial dimension, across the clusters and with respect to the pooled model, and temporal dimension. However, strong links and statistical consistency exist between the aggregate results and those reported for the southern and eastern regions.

The rest of the paper is structured as follows. Section \ref{sec2} introduces the main tools adopted in the paper, that is the the clusterwise regression and the spatially clustered regression, while Section \ref{sec3} discusses the Spatially-clustered spatial autoregressive (SCSAR) model. Section \ref{sec4} introduces the empirical application, discussing the problem, its relevance and data. Section \ref{sec5} discusses the main results, while Section \ref{sec6} concludes with final remarks.

\section{Spatially clustered regression}
\label{sec2}

\noindent Let us consider a set of $N$ $(i=1,\dots,N)$ statistical units and let us define $y_i$ the value of the response variable for the $i$-th statistical unit. Moreover, let $\mathbf{x}_i=[x_{i1}, \dots, x_{iP}]'$ be the $P$-dimensional vector of covariates associated with the $i$-th unit, with $x_{ip}$ the value of the $p$-th covariate for $i$-th unit. The standard linear regression model can be written as
\begin{equation}
\label{eq:reg0}
    y_i = \sum_{p=1}^{P} \theta_p x_{ip} + \varepsilon_i, \quad i=1,\dots,N; p=1,\dots,P,
\end{equation}
where $\theta_p$ is the $p$-th component of the $P$-dimensional vector of regression coefficients $\boldsymbol{\theta}=[\theta_1, \dots, \theta_P]'$ and $\{\varepsilon_i\}$ $i=1,\dots,N$ are i.i.d. error terms. In this section, we provide the theoretical foundations of the spatially clustered regression model. We introduce clusterwise regression first and then discuss its extension to the case with spatial clustering.

\subsection{Clusterwise regression: generalities}


Let us assume the existence of $K$ $(k=1,\dots,K)$ clusters and define $N_k$ the size of the $k$-th cluster. We denote by $y_{ik}$ the value of the response for the $i$-th statistical unit in the $k$-th cluster and by $\mathbf{x}_{ik}=[x_{ik1}, \dots, x_{ikP}]'$ the $P$-dimensional vector of covariates associated with $y_{ik}$. Therefore, model \eqref{eq:reg0} for the $k$-th cluster becomes
\begin{equation}
\label{eq:reg1}
    y_{ik} =  \sum_{p=1}^{P}\theta_{pk} x_{ikp} + \varepsilon_{ik}, \quad i=1,\dots,N_k;k=1,\dots,K
\end{equation}
with $N_k < N$ being number of units belonging to the $k$-th cluster, $\boldsymbol{\theta}_k=[\theta_{1k}, \dots, \theta_{Pk}]'$ the $P$-dimensional vector of unknown regression coefficients of the the $k$-th cluster. Following previous studies, we assume that the conditional density of $y_{ik}$ given $x_{i k}$,  $f_i\left(y_{ik} \mid \mathbf{x}_{ik}; \boldsymbol{\theta}\right)$, is known and, therefore, 
the parameters within each cluster $k$ can be estimated by maximizing the following log-likelihood function
\begin{equation}
\label{eq:func}
    Q(\boldsymbol{\theta},k) = \sum_{i=1}^{N} \log f_{ik} \left(y_i \mid \mathbf{x}_i, \boldsymbol{\theta}_{k} \right),
\end{equation}
given that the clustering assignment of each $i$-th unit is known. However, the clustering structure is not known and a $k$-means-like algorithm is needed for estimating the cluster membership and the within-cluster coefficients iteratively (e.g., see \cite{ari2002clustered,coppi2006least,hathaway1993switching,spath1979algorithmus}). The $k$-means approach works as follows. Considering an initial partition of the $N$ units into $K$ clusters, the parameters of the within-cluster regression are estimated for each $k$-th cluster. Next, the units are re-assigned by maximizing the total likelihood function \eqref{eq:func}. These two steps, estimation and cluster assignment, are iterated until convergence. 


\subsection{Clusterwise regression with spatial constraint}

\noindent We now assume that the $N$ units are georeferenced and that the location of the $i$-th unit, $s_i$, is known. To identify the geolocation of the variables, we write $y_i (s_i)$ and $\mathbf{x}_i (s_i)$. According to the first law of geography \cite{tobler1970computer}, \emph{everything is related to everything else, but near things are more related than distant things}. Therefore, in the regression framework, it is known that neighbor observations may share a similar relationship between dependent variables and covariates. To account for spatially homogeneous regression functions in the case of spatial data, it is therefore important to introduce spatial clustering in the clusterwise regression problem \eqref{eq:reg1}. 

To enhance spatial clustering, the objective function \eqref{eq:func} has to be penalized by increasing the likelihood when contiguous units are clustered together. For the computation of the spatial penalty, we need to introduce the spatial weighting matrix $\mathbf{W}$, which is a square symmetric matrix of order $N$ where the diagonal elements are set to zero, and the generic off-diagonal element $w_{ij}$ provides information on the spatial proximity of two units $i,j$, given that $i \neq j$. Specifically, following \cite{sugasawa2021spatially}, we assume that the elements $w_{ij}$ are binary, that is,
\begin{equation*}
w_{ij}= \begin{cases}
1 & \text { if unit } i \text { is contiguous to unit } j, \\
0 & \text { otherwise.}
\end{cases} 
\end{equation*}
We notice different definitions of $\mathbf{W}$ are possible (e.g., see \cite{elhorst2014spatial}).  In particular,
 \cite{sugasawa2021spatially} proposed the use of the following penalized likelihood
\begin{equation}
\label{eq:func2}
    Q(\boldsymbol{\theta},k) = \sum_{i=1}^{N} \log f_{ik} \left(y_i \mid \mathbf{x}_i, \boldsymbol{\theta}_{k} \right)+\phi \sum_{i<j} w_{ij} I(k_i = k_j),
\end{equation}
the parameter $\phi$ controls for the relevance of spatial penalty,  $I$ is the indicator function and $k_i$ represents the cluster assignment of the $i$-th unit. Therefore, the role of the penalty $\sum_{i<j} w_{ij} I(k_i = k_j)$ is to increase the value of the likelihood function when two contiguous units $i$ and $j$ are in the same cluster. The penalty term in \eqref{eq:func2} is motivated by a spatial process for discrete space known as the \cite{potts1952some} model. The ultimate aim of the so-defined spatially clustered regression model is to maximize the likelihood, by maximizing the \eqref{eq:func2} that enhances spatial clustering. 

However, maximizing the penalized likelihood \eqref{eq:func2} does not ensure that all contiguous units are clustered together. Let us assume that two units $i$ and $j$ are spatially contiguous, i.e. $w_{ij}=1$, but they do not share the same regression coefficients for some reason. In this setting, the algorithm does not force the two units to be in the same clusters. If the likelihood $ f_{ik} \left(y_i \mid \mathbf{x}_i, \boldsymbol{\theta}_{k} \right)$ and $ f_{jk^\prime} \left(y_j \mid \mathbf{x}_j, \boldsymbol{\theta}_{k^\prime} \right)$ are completely different, these units will be assigned in different clusters, $k_i \neq k_j$, even if $i$ and $j$ are contiguous. If the difference is negligible, however, the maximization of the spatially penalized objective function lets the two units be clustered together.

The estimators of $\boldsymbol{\theta}$ and $k$ can be found by maximizing iteratively the \eqref{eq:func} using a $k$-means regression algorithm. Therefore, given an initial partition of the $N$ units into $K$ homogeneous clusters, where $\widehat{k}_i$ represents the assignment of the $i$-th unit in the $k$-th cluster, we can estimate the parameters within the $k$-th cluster in a first step  as follows \cite{sugasawa2021grouped}
\begin{equation}
    \widehat{\boldsymbol{\theta}}_k = \underset{\boldsymbol{\theta}_k}{\arg\max}  \sum_{i=1}^{N} I(\widehat{k}_i =k) \log f_{ik} \left(y_i \mid \mathbf{x}_i, \boldsymbol{\theta}_{k} \right).
\end{equation}
Then, in a second step, the clustering assignment is obtained from the penalized likelihood, that is,
\begin{equation}
    \widehat{k}_i = \underset{k_i \in \{1,\dots,K\}}{\arg\max} \Bigg\{\log f_{ik_i} \left(y_i \mid \mathbf{x}_i, \widehat{\boldsymbol{\theta}}_{k_i}\right)+\phi \sum_{i<j} w_{ij} I(k_i = k_j)\Bigg\}.
\end{equation}
These two steps are alternated until convergence. The final results may depend on the choice of $\phi$ and the initial partition. About the first point, \cite{sugasawa2021grouped} shown by a simulation study that results are not much sensitive to $\phi$ and suggest $\phi=1$. About the second issue, $k$-means clustering based on the Euclidean distance on geographical coordinates is often considered a suitable initial partition. 

\section{Spatially-clustered spatial autoregressive model}
\label{sec3}

\noindent In spatial statistics, the consideration of spatial autocorrelation is essential for accurately modeling relationships between variables across geographical space. Spatial autocorrelation refers to the phenomenon where observations in nearby locations exhibit correlation, violating the independence assumption of traditional regression models \cite{elhorst2014spatial}. Failure to account for spatial autocorrelation can lead to biased parameter estimates and inefficient inference.

Spatially clustered regression models are valuable tools in spatial analysis as they allow us to identify and account for similarities among neighboring units, which can improve the accuracy of the regression results. However, it is important to recognize that by grouping spatially contiguous units into the same clusters, we may inadvertently exacerbate the issue of spatial autocorrelation. To address this concern, it is crucial to incorporate techniques that explicitly account for spatial autocorrelation. This may involve the use of spatial autoregressive models (SAR), spatially correlated error terms (SEM) and/or spatially lagged variables in our regression framework in a spatially clustered linear regression model \eqref{eq:reg1}. By doing so, we can effectively account for the induced spatial autocorrelation resulting from spatial clustering, ensuring that our regression models provide reliable and robust estimates.


In what follows, we focus on the Spatial Autoregressive (SAR) model which accounts for spatial autocorrelation by including a spatial lag of the dependent variable as an additional regressor in the equation \eqref{eq:reg0}, that is
\begin{equation}
\label{eq:sar}
y_i = \rho \sum_{j=1}^{N} w_{ij} y_j + \sum_{p=1}^{P} \theta_{p} x_{ip} + \varepsilon_{i},
\end{equation}
where $\rho$ is the spatial autoregressive coefficient. The spatial lag term captures the spatial effect, indicating that the value of the dependent variable at a given location is influenced by the values of the dependent variable in the neighboring area.

The parameters in model \eqref{eq:sar} can be estimated with maximum likelihood. By denoting $S_N (\rho) = I_N - \rho \mathbf{W}$, with $I_N$ the $N$-dimensional Identity matrix and under the Gaussian assumption, the log-likelihood is \cite{lee2004asymptotic}
\begin{equation}
\label{eq:sarloglik}
   \begin{aligned}
\log f (y_i \mid  x_i, \mathbf{W}, \boldsymbol{\delta}) = -\frac{N}{2} \ln (2 \pi) -\frac{N}{2} \ln \sigma^2 + \ln \left|S_N(\rho)\right| \\
-\frac{1}{2 \sigma^2} \sum_{i=1}^{N} \left( y_i - \rho \sum_{j=1}^{N} w_{ij} y_j - \sum_{p=1}^{P} \theta_{p} x_{ip} \right)^2.
\end{aligned}
\end{equation}
with $\boldsymbol{\delta}=[\rho, \boldsymbol{\theta}]'$ and $\sigma^2$ the unknown residual variance. Let us consider the spatially clustered extension of the SAR model \eqref{eq:sar}. We generalize the algorithm of \cite{sugasawa2021spatially} to this setting, by estimating the following equations
\begin{equation} \label{eq:scsar}
  y_{i,k} = \rho_k \sum_{j=1}^{N} w_{i_k j_k} y_{j_k} + \sum_{p=1}^{P} \theta_{p,k} x_{ip,k} + \varepsilon_{i,k}, \quad \forall k=1,\dots,K
\end{equation}
where $w_{i_kj_k}$ are the (binary) weights associated with the spatial weighting matrix $\mathbf{W}$ for the units belonging to the $k$-th cluster. These quantities provide information on the spatial contiguity of the units in the $k$-th cluster.

The objective function to maximize
\begin{equation}
\label{eq:func3}
    Q(\boldsymbol{\delta},k) = \sum_{i=1}^{N} \log f_{ik} \left(y_i \mid \mathbf{x}_i, \mathbf{W},  \boldsymbol{\delta}_{k}\right)+\phi \sum_{i<j} w_{ij} I(k_i = k_j),
\end{equation}
with $\boldsymbol{\delta}_{k}=[\rho_k, \boldsymbol{\theta}_{k}]$' be the vector of coefficients estimated for the $k$-th cluster and $\mathbf{W}$ is the full spatial weighting matrix. Like \cite{sugasawa2021spatially}, we introduce the spatial penalty term to the SAR log-likelihood for enchaining the spatial clustering.


Given the obtained cluster structure, the model parameters within each cluster are obtained by maximizing the likelihood, that is,
\begin{equation}
      \widehat{\boldsymbol{\delta}}_k = \underset{\boldsymbol{\delta}_k}{\arg\max}  \sum_{i=1}^{N} I(\widehat{k}_i =k) \log f_{ik} \left(y_i \mid \mathbf{x}_i, \mathbf{W}, \boldsymbol{\delta}_k \right).
\end{equation}
By using $I(k_i=k)$ we restrict $\mathbf{W}$, so that $\rho_k$ and $\boldsymbol{\theta}_k$ for each clusterwise regression are obtained on restricted versions of the observed sample.
Then, the clustering rule depends on the estimated parameters, $\widehat{\boldsymbol{\delta}}_k$ $\forall k=1,\dots,K$, and it equals to
\begin{equation}
    \widehat{k}_i = \underset{k_i \in \{1,\dots,K\}}{\arg\max} \Bigg\{\log f_{ik} \left(y_i \mid \mathbf{x}_i, w_{ij},  \widehat{\boldsymbol{\delta}}_{k} \right)+\phi \sum_{i<j} w_{ij} I(k_i = k_j)\Bigg\},
\end{equation}
which maximizes the objective function at each step. The algorithm is shown in detail in Table 1.

\begin{algorithm}
  \SetAlgoLined
  \SetKwInOut{Input}{Input}
  \SetKwInOut{Output}{Output}

  \Input{Response variable $(\mathbf{y})$, Covariates $(\mathbf{X})$, Spatial weights matrix $(\mathbf{W})$, Number of clusters ($K$), Penalty $(\phi)$, Maximum iterations ($\text{max.itr}$), Tolerance of the log-likelihood $(\eta)$.}
  \Output{Final cluster assignments, Clustered regression results}

  \BlankLine
  Initialize cluster assignments $\mathbf{k}^{r=0}$ (e.g., $k$-means based on spatial coordinates);\;

  \While{$\mathbf{k}^{r} \neq \mathbf{k}^{r-1} \text{ and } r\leq\text{max.itr} \text{ and } |(\ell^{r}-\ell^{r-1})/\ell^{r-1}| > \eta$}{
    \tcp{Update step A: group-wise parameters}
    \For{k in 1:K}{
      $\widehat{\boldsymbol{\delta}}^{r}_k = \underset{\boldsymbol{\delta}_k}{\arg\max} \sum_{i=1}^{N} I(\widehat{k}^{r-1}_i=k) \log f_{ik} \left(y_i \mid \mathbf{x}_i, \mathbf{W}, \boldsymbol{\delta}_k \right)$;\;
    }


    \tcp{Update step B: membership}
    \For{i in 1:N}{
      $\widehat{k}^r_i = \underset{k_i \in \{1,\dots,K\}}{\arg\max} \Bigg\{\log f_{ik} \left(y_i \mid \mathbf{x}_i, \mathbf{W},  \boldsymbol{\delta}_{k^{r-1}} \right) + \phi \sum_{j=1, i\neq j}^{N} w_{ij} I(k = \hat{k}^{r-1}_j)\Bigg\}$;\;
    }
    
    $r \leftarrow r+1$\;
  }
  \caption{Spatially Clustered SAR (SCSAR)}
\end{algorithm}


\noindent \textbf{Remark.} We remark that other specifications for spatial regression models are possible. For instance, the spatial model with lagged predictors (SLX) and the spatial error model (SEM). The SCSAR algorithm is exactly the same for these two specifications, while the penalized likelihood is different. For the sake of brevity, we discuss with more details these two possible extensions in the Appendix \ref{appendix1}.

\section{Empirical analysis of the statistical concentration in the European agricultural market} \label{Sec_Application}
\label{sec4}
he SCSAR model introduced in the Section \ref{sec3} is now applied to a real-world dataset related to the agricultural market in Europe. Specifically, we aim at investigating the spatio-temporal dynamic of a measure of statistical concentration on  production (standard output) in the European agricultural industry at the regional level. Here, spatial autoregressive models are employed to assess the presence of spatial spillovers of market concentration across regions assuming that the effects are local at not global. That is, both the spatial spillovers and the magnitudes (as well as their significance) of the covariates on the observed concentration index are heterogeneous across unknown spatial clusters of regions.

\subsection{Concentration of the Agricultural market: macro-trends in Europe and in the World}

On a worldwide basis, agriculture is still one of the largest production sectors on which the world's economies rely. According to World Bank estimates, in 2022 the share of GDP generated by the agricultural sector is 4.1\% \cite{WB_AgroVA}, with year-to-year growth rates steadily above 3\% since 1980 \cite{WB_AgroVAg}, and involving several million people on a daily basis (in 2022 26\% of the world's workers are employed in the agricultural sector, but in 1991 it was 44\% as reported in \cite{WB_AgroEmp}). The number of active farms is growing and ranges between 570 million \cite{LowderEtAl2016} and 610 million \cite{LowderEtAl2021}. Of these, family farms account for between 80\% and 90\% of the total \cite{FAO2014} and produce about 80\% of the global food in value terms \cite{RicciardiEtAl2018}. According to \cite{LowderEtAl2021} estimates, 84\% of global farms are small farms but occupy only 12\% of all agricultural land produce 1/3 of the world's food. Global production is in fact under the control of a very small number of huge farms such that 1\% of the largest farms cover more than 70\% of the available farmland \cite{LowderEtAl2021}, with a sharply increasing trend from the previous decade \cite{LowderEtAl2016}. This evidence is clearly indicative of a growing process of concentration in the global agricultural market that favors large-scale producers at the expense of small-scale farming and with an increasing environmental footprint.

The same trends are also confirmed for Europe. \cite{LowderEtAl2021} observe that in the European countries with the largest agricultural area (i.e., France, Germany, Spain, and the UK), in terms of farmland distribution between 2005 and 2013, in none of the cases is there an increase in the share of small farms; while with the sole exception of Spain, the share of agricultural area managed by farms larger than 100 hectares increased. Similar results are also noticeable for Brazil and the United States of America. Less recently, \cite{Blacksell2010} estimated that in Bulgaria, Czech Republic and Hungary the top 10\% of largest farms operate on the 80\% of the utilized agricultural land, whereas in Slovenia, Poland, Romania and Estonia, the same share of farms covers only 40\% of the agricultural area.

Despite the EU adopted a Common Agricultural Policy (CAP) decades ago \cite{Stead2007}, differences among countries in terms of management of the internal agricultural markets and land-management regimes are well established in the literature \cite{JepsenEtAl2015}. Often, Europe is represented as a fragmented framework with marked heterogeneity in the productive specializations (for instance, the co-existence of crops and livestock in the same area), in farming styles (such as the peasantry based on agrarian lifestyles, small size and weak dependence on the market discussed in \cite{vanderPloegEtAl2009}) and economic capacity and power \cite{GuarinEtAl2020}. Also, considerable differences concerning the farmers’ backgrounds and development paths \cite{LowderEtAl2016} and the production assets owned by the farms based on their size (see, for instance, the discussion on heterogeneity regarding the presence of subsistence farming in the EU by \cite{DavidovaEtAl2012}) can be detected. Such a fragmentation in the agricultural industry directly reflects in a pronounced multifaced territorial (spatial) heterogeneity \cite{JepsenEtAl2015} testified by a considerable variability within and between regions and countries with respect to the farms’ structural characteristics \cite{GuiomarEtAl2018}. Specifically, \cite{GuiomarEtAl2018} classified the European provinces according to the relevance of small farms in the agricultural and territorial context of each region and considering different dimensions of farm size. Their findings suggest that Europe can be framed into several subareas with different levels of relevance of the agricultural activities based on the combination of structural (physical) size and economic size of farms. For instance, France and Belgium, together with provinces in the northern Germany, southern and eastern Ireland, south-western and eastern Scotland, The Netherlands and Denmark are characterized by a very low proportion of small-scale farms and by a high share of large and high-income farms. Conversely, farms with small economic size are common in Hungary, southern Slovakia, northern and eastern Poland and in southern Austrian regions. Also, spatial heterogeneity appears to be a relevant element when exploring the productivity of the agricultural sector in Europe (see, for instance, \cite{GiannakisBruggeman2018} regarding the spatial determinants of agricultural labor productivity or \cite{Martínez-VictoriaEtAl2018,Martínez-VictoriaEtAl2019} on the productivity growth on spanish agri-food firms).

Recalling \cite[ Section 2.3]{ZhuEtAl2018}, statistical modeling of data in spatial contexts requires that relationships between variables remain static throughout the area under consideration. However, this seems unrealistic "considering that geographic phenomena are inherently heterogeneous," especially when "considering large and complex geographic areas". Spatially-clustered regression was expressly developed as a statistical tool that can address the problem of spatial heterogeneity, in contexts where the relationships between variables change in a given space \cite{sugasawa2021spatially}. In our case, we use data from over 200 regions on the European continent, covering an area of approximately 4045 millions of squared kilometers and administered by 24 countries with common rules and common market, but with the significant structural differences and local characteristics discussed in the previous section. Therefore, we believe it is reasonable to assume that the explanatory variables considered may generate spatially uneven effects on the observed market concentration at the regional level. Hence the need to compare potential aggregate effects (estimated by a pooled regression model) common to all European regions with the potential local effects in subgroups of regions with commonalities in their agricultural structure (estimated by local regression models).

\subsection{Regional data on the agricultural market for Europe}
Recall that the EU adopts an administrative classification known as the Nomenclature of territorial units for statistics classification or NUTS \cite{EurostatNUTS}, which subdivides country members into a multi-layer hierarchical structure having as the lower stratum the local are units or municipalities (LAUs), in turn uniquely nested within provinces (NUTS-3), regions (NUTS-2), macro-regions (NUTS-1), and countries (NUTS-0). In the present work, we use the regional 2010 NUTS-2 classification as it allowed for the largest spatial and temporal coverage of the agricultural industry data. Notice that, despite more recent classification are available, we adopted the NUTS 2010 nomenclature as it is the only one fully covering the 27 countries currently belonging to the European Union within the 2010-2020 period. Also, we preserve only the European regions having complete (i.e., non-missing) information for the entire time span and we removed from the analysis all the non-continental regions and the islands. According to the former, we exclude from the analysis all the extra-European administrative units, such as Guyane, Martinique, and Réunion for France or Ciudad de Melilla for Spain; while according to the latter, we removed data on counties that are islands, that is, Cyprus, Ireland and Malta. We made this choice to have a specific focus on continental Europe. Indeed, while extra continental units would be clustered together due to the dominance of the geographical distance with respect to the features' distance, the other three major islands would introduce the issue of dealing with spatial units without neighbors or with a few number of neighbors that belong to the same country. The selected dataset includes complete data for 221 NUTS-2 regions across Europe (see Figure \ref{fig:cluster2010} or Figure \ref{fig:cluster2020} for a graphical representation of the available spatial units). 

We consider a regionalized dataset built by combining several databases from Eurostat covering the period from 2010 to 2020. Relevant agricultural market information were collected from the “Main farm indicators by NUTS 2 regions” open database \cite{EurostatAgroData}, which provides several structural and economic indicators on the agricultural industry in Europe for the years 2010, 2013, 2016, and 2020. The database contains, among others, data on the overall number of agricultural holdings or farms (measures as count), the utilized agricultural land (measured in hectares), and the standard output from agricultural production (measured in Euros). Notice that, despite data are available for several subclassification of farms with respect to the productive specialization, such as the organic producers or the livestock-specialized farms, our focus is on the overall European agricultural industry. Therefore, we considered altogether the farms regardless their business specificities. Future extensions of the current project could take into account such information and extend the market concentration analysis by characterizing the findings with respect to the farms’ typologies.

In the Eurostat database, economic values of agricultural production are stratified into eleven classes of economic size with increasing values of the standard output. In Table \ref{Tab_FarmsEU28} we report the number of farms (thousands) by classes of economic size for the whole European Union (28 countries) from 2010 to 2020. 
\begin{table}[ht]
\centering
\caption{Number of farm holdings (thousands) for the whole European Union (28 countries) by economic size classes. The column $\Delta_{2020-2010}$ reports the observed raw variation from 2010 to 2020.} 
\label{Tab_FarmsEU28}
\begin{tabular}{lccccc}
  \hline
  Standard output & 2010 & 2013 & 2016 & 2020 & $\Delta_{2020-2010}$\\ 
  \hline
Zero euros &  1029 &   895 &   541 &   242 &  -787 \\ 
Over zero euros to less than 2000 euros &  7145 &  5686 &  4809 &  4487 & -2658 \\ 
From 2 000 to 3 999 euros &  6411 &  5524 &  4897 &  4369 & -2042 \\ 
From 4 000 to 7 999 euros &  9501 &  8509 &  7294 &  6551 & -2950 \\ 
From 8 000 to 14 999 euros & 11388 & 10263 &  9087 &  8134 & -3254 \\ 
From 15 000 to 24 999 euros & 11336 & 10285 &  9379 &  8607 & -2729 \\ 
From 25 000 to 49 999 euros & 19616 & 18212 & 16447 & 15702 & -3914 \\ 
From 50 000 to 99 999 euros & 25818 & 24726 & 22303 & 21058 & -4760 \\ 
From 100 000 to 249 999 euros & 37495 & 37542 & 36289 & 33920 & -3575 \\ 
From 250 000 to 499 999 euros & 20429 & 22972 & 24162 & 22583 &  2154 \\ 
500 000 euros or over & 25678 & 29676 & 30404 & 29441 &  3763 \\ 
   \hline
\end{tabular}
\end{table}
The table clearly reveals that between 2010 and 2020 there has been a process of major reduction in the number of small agricultural producers (farms with standard output of less than 8000 has been reduced by almost 8 million) and at the same time a large increase in farms with large productions. For example, in ten years farms with production over 250 thousand € has increased by 5.9 million, about +12\% in the decade. Medium-sized farms have also declined significantly over the decade, with negative variations of between three and four million for each economic size group. This preliminary evidence is consistent with previous research, such as \cite{LowderEtAl2016}, which estimated that between 1960 and 2000, that average farm size decreased in low- and lower-middle-income countries, while it increased in upper-middle-income and high-income countries (most of the European countries fall in the latter category). 

Such a shift of production into the hand of a limited number of very large companies, at the expense of small and medium-sized producers, legitimizes the suspicion of a significant increase in the concentration of the European agricultural market.

To quantify the agricultural market concentration, we compute the Gini index \cite{Gini1921} on the standard output (production) owned by farm holdings grouped according to the Eurostat classification. The Gini index over production can be interpreted as a statistical measure of concentration with respect to the economic capacity of farms; indeed, by considering a 0-100 range, a value of the index close to 100 suggests a high market concentration in which the production is owned by a restricted number of companies (or by a single farm in the case of Gini exactly one), while value close to 0 means that production is equally-distributed across the farms and that the concentration is weak (for a recent and in-depth review about the statistical properties and the interpretation of the Gini index see \cite{GiorgiGigliarano2017}).

The Gini index is computed using the \cite{Brown1994} specification for grouped data. Specifically, we leverage on the stratification of agricultural farms according to their economic size and we compute, for each region $s=1,\ldots,N=221$, the Gini index for the standard output as follows:
\begin{equation}
Gini_{s} = \frac{N_{s}}{N_{s}-1} \left[ 1 - \sum_{j=1}^{J}{\left(Q_{j}+Q_{j-1}\right) \cdot \left(F_{j}-F_{j-1}\right)} \right] 
\end{equation}
where $N_{s}$ is the total number of farms in region $s$; $Q_{j} = \sum_{i=1}^{j}{q_{j}}$ is the cumulative proportion of production (cumulated standard output) up to the $i$-th ordered class (with $q_{i}$ being the share of production over the total associated with the $i$-th ordered class), and $F_{j} = \sum_{i=1}^{j}{f_{i}}$ is the cumulative proportion of farm holdings up to the $i$-th ordered class (with $f_{i}=\frac{N_{si}}{N_s}$ being the share of production over the total associated with the $i$-th ordered class).

From Eurostat we collected additional features on both macroeconomic, geographical, and agricultural-related economic dimensions that could help in explaining both the observed values for the Gini coefficient in Europe and the variation occurred during the decade from 2010 to 2020. To model the spatialized Gini index for the standard output (response variable), we consider a set of 10 covariates referring to four dimensions: regional wealth, regional labor market, regional economy, and regional landscape and environment. In Table \ref{Tab_Xmat}, we report a description of the selected variables used in the clustered regression exercise.

\begin{table}
\caption{\centering{Regional (NUTS-2) variables from Eurostat selected as relevant covariates for the spatially clustered-regression on Gini for production}}
\label{Tab_Xmat}
\setlength\extrarowheight{10pt}
\begin{center}
\begin{tabular}{|M{3cm}|M{3.5cm}|M{3.5cm}|M{2.5cm}|}
\hline
\textbf{Dimension} & \textbf{Description} & \textbf{Short name} & \textbf{Unit of measure} \\
\hline
Regional wealth             & Per capita GDP   & GDP pc   & €/person  \\
\hline
Regional labor market       & Share of employed persons in NACE sector A over total employment  & Share of agro employment & \% of total employed persons       \\
Regional labor market       & Worked hours per worker of the NACE sector A & Worked hours per agro worker & Hours/worker \\
\hline
Regional economy            & Gross Value Added (GVA) of NACE sector A per agricultural worker & GVA per agro worker  & Ths€/worker \\
Regional economy            & Gross Fixed Capital Formation (GFCF) of NACE sector A per agricultural worker & GFCF per agro worker  & Ths€/worker \\
Regional economy            & Share of GVA from NACE sector A over total GVA & Share of agro GVA & \% of total GVA \\
\hline
Regional landscape and environment & Share of agricultural land over total land & Share of agro land & \% of total land \\
Regional landscape and environment & Average altitude    & Average altitude & meters       \\
Regional landscape and environment & Heating Degree Days & HDD & Degrees       \\
\hline
\multicolumn{4}{M{14cm}}{\scriptsize{NACE refers to the classification of economic activities in the European Union revision 2 update 1 (NACE rev. 2.1). Specifically, sector A includes agriculture, forestry and fishing. For details, see the dedicated webpage at \href{https://ec.europa.eu/eurostat/web/nace}{https://ec.europa.eu/eurostat/web/nace}}}
\end{tabular}
\end{center}
\end{table}

The list includes both generalist variables i.e., those capturing factors relevant not only for agriculture, but for society as a whole (e.g., GDP per capita), but also variables specifically referring to farming activities. Among the latter, we have enclosed variables referring to the relevance of agricultural industry on the regional labor market (share of employment in agriculture) and on the regional economy (e.g., share of agricultural GVA on total GVA); variables measuring the intensity of production and profit relative to the number of people employed in the sector (e.g., hours worked per agro-employed and GVA per agro-employed); and variables measuring the propensity to invest in the agricultural sector (e.g., GFCF per agro worker). Finally, to control with respect to the geographic characteristics of the territories, we included proxies of temperature (e.g., Heating Degree Days) and landscape and and local orography (e.g., share of agricultural land and average altitude). Notice that, several variables refer to the EU-NACE classification \cite{NACE} and, specifically, make use of data concerning the A sector (i.e., agriculture, forestry and fishing). Although this is a somewhat less fine aggregation than the required one, this is perhaps the best possible approximation for macroeconomic and industrial data at the regional level provided by Eurostat. It is noteworthy that, on the one hand, we expect that variables such as regional wealth and environment do not directly impact agricultural market concentration as they involve more complex socio-economic and contextual dynamics; on the other hand, the economic relevance of the agricultural supply chain, labor market, and investment may have generated effects (both positive and negative) on agricultural industrial dynamics by favoring or counteracting the concentration process. Moreover, these effects are expected to be non-uniform across European regions.

\section{Empirical results}
\label{sec5}

\begin{figure}[!htb]
    \centering
    \includegraphics[width=\linewidth]{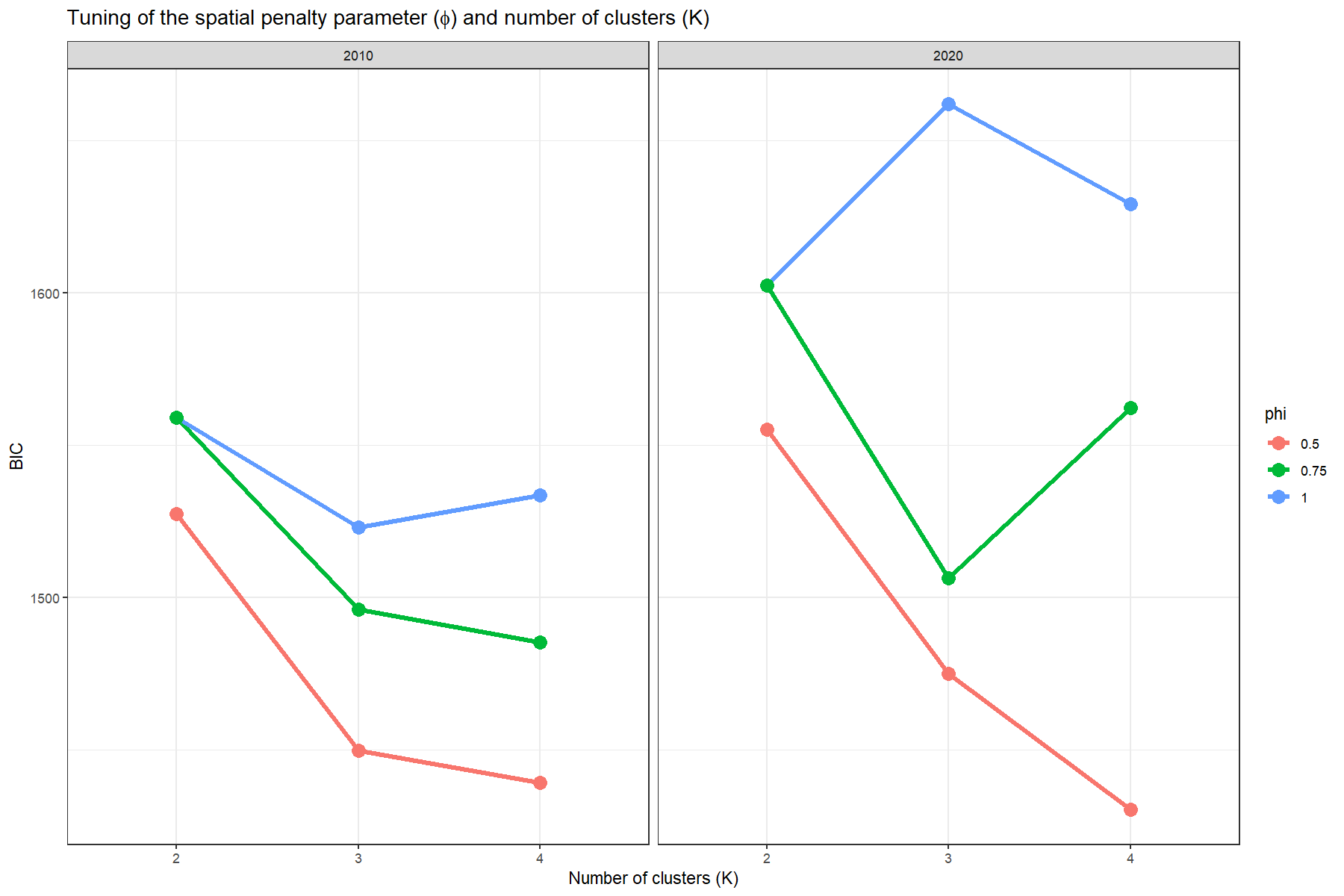}
    \caption{Tuning the parameters $K$ and $\phi$ of the SCSAR model, considering the sets $K=\{2,3,4\}$ and $\phi=\{0.5, 0.75, 1\}$. The solution is given by the Elbow for the solution minimizing the BIC criterion.}
    \label{fig:tuning}
\end{figure}

In what follows we discuss the results obtained with the real data, considering the estimation of SAR-like models with exogenous covariates as in \eqref{eq:sar} for the years 2010 and 2020. We compare, evaluating the differences between the two years, the results of the SAR model without clustering with those obtained with the SCSAR discussed in Section \ref{sec3}. Before applying the SCSAR model, we need to tune two parameters, that is, the amount of spatial penalty $\phi$ and the number of clusters $K$. In accordance with previous studies (e.g. see, \cite{zhao2008knee}, we tune the parameters with the aim of minimizing the BIC criterion while keeping the number of clusters parsimonious. Indeed, the sample size of European NUTS is $N=222$ and a distinction in many clusters may lead in too small cluster sizes $N_k$, for some $k$, which would lead unreliable inference. Therefore, we choose in the set $K=\{2,3,4\}$ and $\phi=\{0.5, 0.75, 1\}$ with the Elbow method for the solution with minimal BIC criterion \cite{sugasawa2021spatially}. According to Figure \ref{fig:tuning}, which provides a summary of the different results, we choose $\phi=0.5$ and $K=3$ for both the years 2010 and 2020. Indeed, the lowest BIC values are achieved for both the years when $\phi=0.5$ and the Elbow is detected for $K=3$. 

\begin{figure}[!htb]
    \noindent\makebox[\textwidth]{\includegraphics[width=1.3\linewidth]{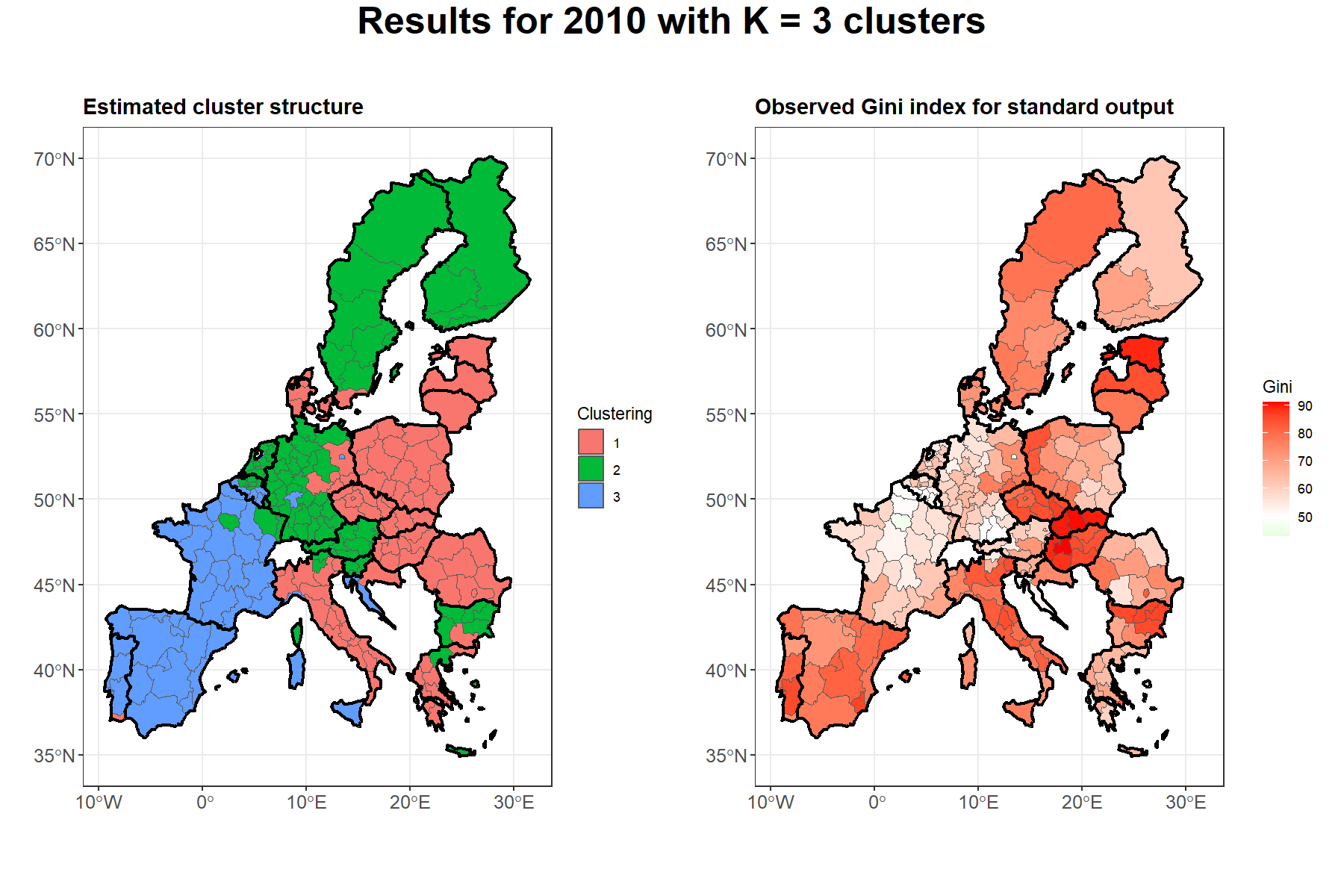}}
    \caption{Estimated clustering structure for 2020 (left panel) and observed Gini index for standard output (right panel) in 2010.}
    \label{fig:cluster2010}
\end{figure}

\begin{figure}[!htb]
    \noindent\makebox[\textwidth]{\includegraphics[width=1.3\linewidth]{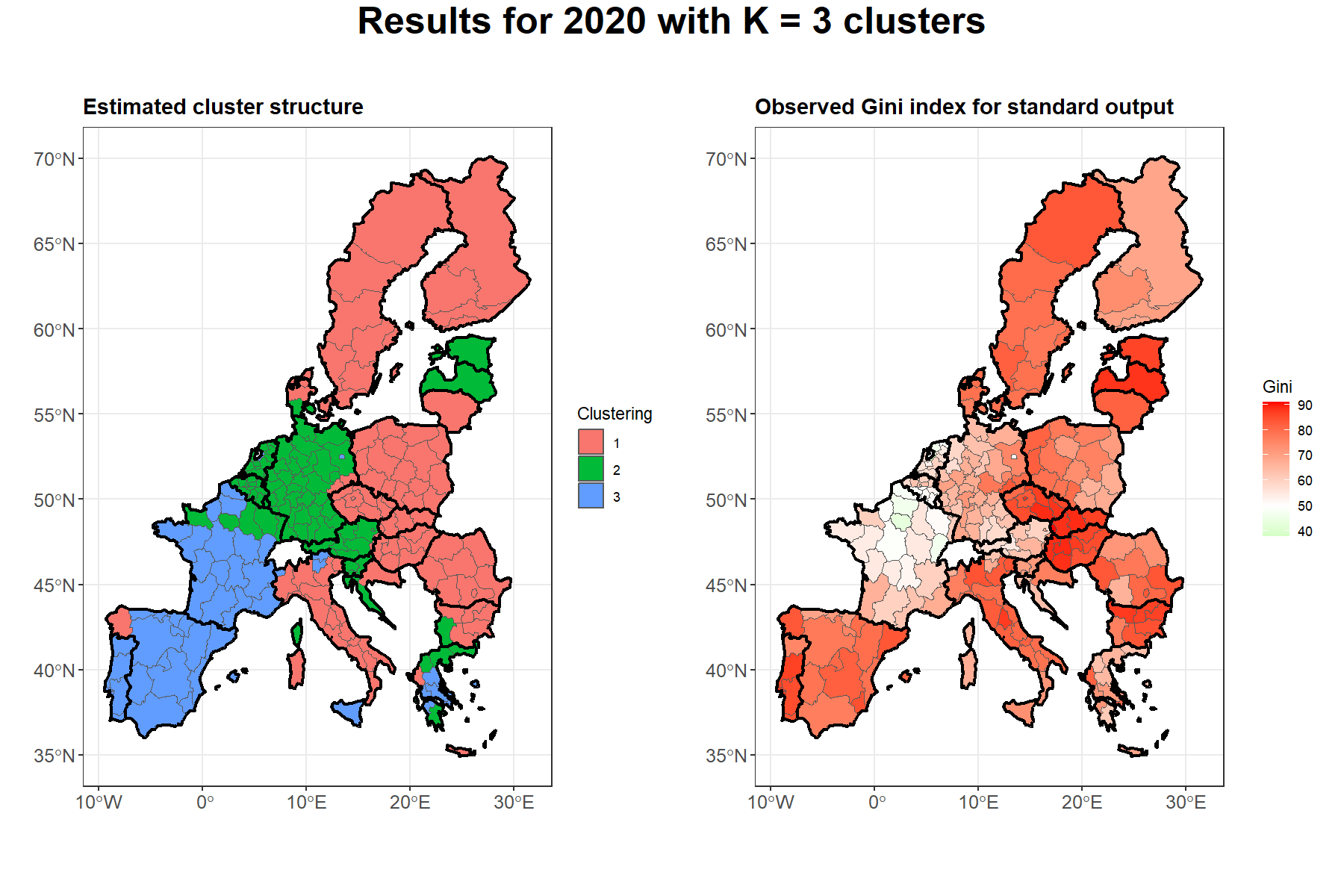}}
    \caption{Estimated clustering structure for 2020 (left panel) and observed Gini index for standard output (right panel) in 2020.}
    \label{fig:cluster2020}
\end{figure}

Table \ref{table:coefficients} provides a comprehensive overview of the main results, distinguishing between the years 2010 and 2020 and showing the parameter estimates, the standard errors and the LR test for both the SAR model with the full-sample (pooled model in Table \ref{table:coefficients}) and the SCASAR model for each cluster. The clusters are shown in Figure \ref{fig:cluster2010}, for the year 2010, and in Figure \ref{fig:cluster2020}, for the year 2020.

Let us discuss first the cluster structures find by the SCSAR algorithm. For the year 2010 (Figure \ref{fig:cluster2010}), we find the existence of two equally sized clusters (Cluster 1 and Cluster 3) and a smaller one (Cluster 2). However, albeit Cluster 1 and Cluster 3 include an equal number of NUTS aggregates, we notice that Cluster 3 mainly includes Spain and France NUTS, while in Cluster 1 we find NUTS from different European countries. In particular, the Cluster 1 includes many NUTS of Italy, Greece, Romania, Hungary, Czech Republic, Slovakia, Poland, Denmark and Baltic countries. Finally, Cluster 3 includes mainly Sweden, Norway, Austria, Germany, most of Benelux and, interestingly, most of Bulgarian NUTS. Figure \ref{fig:cluster2010} also shows, in the right panel, the spatial heterogeneity for the dependent variable. Overall, the NUTS placed in Cluster 1 are those with the highest level of concentration in Europe, while in the other two clusters we generally find NUTS with lower level of concentration. Nevertheless, we notice that the resulting clusters are driven by the existence of spatial spillovers and by the relationship that the observed exogenous variables have with the dependent variable.

Figure \ref{fig:cluster2020} shows the results after 10 years and it is interesting to highlight that the clusters' composition did not change much, albeit some differences can still be found. We can still find a largely sized cluster, that is the Cluster 1, with most of European NUTS in the East of Europe. Moreover, we again find Spain and France NUTS to be clustered together, with Germany clustered with Austria and Benelux countries. Interestingly, compared to the 2010, in the 2020 we find the Nordic countries (Sweden and Norway) in Cluster 1, as well as Bulgaria moving from Cluster 3 to Cluster 1, while Estonia and Latvia moved from Cluster 1 to Cluster 2. The right panel of Figure \ref{fig:cluster2020} indeed highlights an increased market concentration for both Nordic countries and Bulgaria compared to the 2010 and this might partially explain such differences. 

To understand the differences in the cluster structures, however, it is important to thorough analyze the relationships within each cluster. Indeed, we are interested in understanding if some clusters exhibit different levels of spatial autocorrelation and if the determinants of market concentrations differ across the clusters. Moreover, we also compare the pooled model without clustering with the SCSAR results in order to get insight about the relevance of accounting for the cluster structure. We therefore now analyze in detail the results shown in Table \ref{table:coefficients}.

First, we consider the effect of spatial autocorrelation, that is, the $\rho$ coefficient. For the pooled model we find it to be weakly significant in both 2010 and 2020, meaning that the presence of neighbors with high level of market concentration is associated with higher concentration levels in Europe. However, while analyzing the effects at the cluster level, we find our that this spatial effect is statistically significant for Cluster 1 (i.e., Scandinavia and Eastern Europe, Balkans and Italy) but not for the other two clusters. Moreover, considering the results for 2020, we notice a stronger spatial autocorrelation effect for the Cluster 1, which grow from 1\% up to 2\%. Therefore, the pooled model provides an averaged-like evidence and does not highlight such a difference for the clusters.

In terms of covariates impact, we also highlight that the clusters are characterized by different relationships with the dependent variable. Table \ref{table:coefficients} shows the estimates of the coefficients in the pooled regression and clusterwise for the two years, allowing for a direct comparison across space and time. Generally speaking, the coefficients suggest a prominent spatial heterogeneity across the European regions, but also a marked temporal dynamic, leading to an overall favourable attitude toward the clustered strategy. Indeed, in many cases, the coefficients are clearly unstable regarding the magnitudes and the signs (i.e., for several variables, both pooling and groupwise, the signs move from positive to negative or the other way around). A glaring example of this situation is the regional wealth. Globally, we find that the estimated coefficients are poorly statistically significant and that regional wealth was negatively associated with market concentration in 2010, meaning that richer regions were associated with lower market concentration, becoming positive (but not significant) in recent years. A close pattern is found only for for Cluster 3 (i.e., France and Spain), in which the coefficients are strongly significant across time but the sign moves from negative to positive.

We also identify two distinct channels associated with lower market concentrations, that is, the relevance of the agricultural industry in physical terms (i.e., the share of land devoted to agriculture) from one side, and the social relevance (i.e., the share of agricultural employment w.r.t. the total employment) from the other side. According to the pooled model, both the share of employment and the share of agricultural land are globally significant and negatively related to the Gini index. However, the cluster-specific analysis highlights several differences. As for wealth, only Cluster 1 seems consistent with these patterns. Indeed, both Cluster 2 (i.e., Mittel-European regions) and Cluster 3 (i.e., France and Spain) show a positive and statistically significant correlation between concentration and employment (only in 2010) and negative but not significant correlation with agricultural land (for group 3 the share of land has a negative effect in 2010 but loses relevance in 2020). Moving to the economic relevance of the agricultural sector (i.e., the share of GVA from agriculture), we notice that major profits are associated with higher market concentration at the pooled level and for the French-Spanish subarea (in 2020 the coefficient is positive and statistically significant). For the other two clusters, the effect is negative but weakly significant. Lastly, regarding the investments channel (i.e., GFCF per employed person), we find it to be statistically not important in 2010 and have negative inverse signs, albeit it is weakly significant for Cluster 1 and Cluster 2. However, in 2020, the effects become relevant. 


These findings provide several insights regarding the dynamics of the European agricultural market. First, they suggest that using labour force or the utilized land as a proxy of the size of the agricultural market, the higher the socio-territorial relevance of agriculture, the lower the production is concentrated in few and large companies. On the contrary, the greater the value added, the greater the market power of large farms. Second, investments in the agricultural sector gained great importance and became capable of determining market power and business strategy in the European context. Third, positive spatial spillovers (i.e., positive spatial autocorrelation) are relevant only for part of the European countries (actually, those with higher Gini index, i.e. Cluster 1) and at the aggregate level, while market structure and concentration in Western and Central Europe seem to be heterogeneously guided by local drivers.





\color{black}
\begin{sidewaystable}[htbp]
\begin{center}
\caption{Estimated pooled and cluster-wise regression coefficients of SCSAR model for 2010 (columns 2 to 5) and 2020 (columns 6 to 9) using K = 3 clusters and spatial penalty $\phi = 0.50$.}
\begin{tabular}{l | c c c c | c c c c}
\hline
& \multicolumn{4}{c}{Year = 2010} & \multicolumn{4}{c}{Year = 2020} \\
\hline
 & Pooled & K=1 & K=2 & K=3 & Pooled & K=1 & K=2 & K=3 \\
\hline
Intercept                    & $91.10^{***}$ & $78.47^{***}$ & $60.97^{***}$ & $77.31^{***}$ & $83.75^{***}$ & $96.18^{***}$ & $76.32^{***}$ & $82.83^{***}$ \\
                             & $(6.30)$      & $(5.69)$      & $(5.76)$      & $(5.83)$      & $(6.91)$      & $(6.00)$      & $(7.62)$      & $(6.26)$      \\
GDP pc                       & $-0.00^{*}$   & $-0.00$       & $0.00^{***}$  & $-0.00^{***}$ & $0.00$        & $0.00^{*}$    & $-0.00$       & $0.00^{***}$  \\
                             & $(0.00)$      & $(0.00)$      & $(0.00)$      & $(0.00)$      & $(0.00)$      & $(0.00)$      & $(0.00)$      & $(0.00)$      \\
Share of agro employment     & $-1.06^{***}$ & $-0.84^{***}$ & $0.89^{***}$  & $1.05^{***}$  & $-0.47^{*}$   & $-0.66^{***}$ & $0.24$        & $0.17$        \\
                             & $(0.17)$      & $(0.11)$      & $(0.27)$      & $(0.30)$      & $(0.20)$      & $(0.11)$      & $(0.37)$      & $(0.34)$      \\
Worked hours per agro worker & $-0.00$       & $0.00$        & $-0.00$       & $-0.00$       & $-0.00$       & $-0.01^{***}$ & $-0.01^{***}$ & $-0.00$       \\
                             & $(0.00)$      & $(0.00)$      & $(0.00)$      & $(0.00)$      & $(0.00)$      & $(0.00)$      & $(0.00)$      & $(0.00)$      \\
GVA per agro worker          & $-0.14^{*}$   & $0.11$        & $0.25^{**}$   & $0.13^{**}$   & $-0.02$       & $-0.11^{*}$   & $-0.03$       & $0.08$        \\
                             & $(0.06)$      & $(0.09)$      & $(0.08)$      & $(0.04)$      & $(0.06)$      & $(0.05)$      & $(0.06)$      & $(0.06)$      \\
GFCF per agro worker         & $-0.24$       & $-0.17$       & $-0.36^{*}$   & $0.22^{*}$    & $-0.62^{***}$ & $0.42^{***}$  & $-0.37^{*}$   & $-0.16$       \\
                             & $(0.15)$      & $(0.21)$      & $(0.18)$      & $(0.09)$      & $(0.16)$      & $(0.12)$      & $(0.15)$      & $(0.14)$      \\
Share of agro GVA            & $2.49^{***}$  & $1.03^{*}$    & $-0.97^{*}$   & $-1.71^{*}$   & $1.21^{**}$   & $1.62^{***}$  & $-0.52$       & $0.32$        \\
                             & $(0.55)$      & $(0.45)$      & $(0.49)$      & $(0.71)$      & $(0.46)$      & $(0.34)$      & $(0.72)$      & $(0.52)$      \\
Share of agro land           & $-0.17^{***}$ & $-0.19^{***}$ & $0.01$        & $-0.21^{***}$ & $-0.17^{***}$ & $0.02$        & $-0.11^{*}$   & $-0.09$       \\
                             & $(0.04)$      & $(0.04)$      & $(0.03)$      & $(0.04)$      & $(0.05)$      & $(0.04)$      & $(0.05)$      & $(0.05)$      \\
Average altitude             & $-0.00$       & $-0.00$       & $0.01^{***}$  & $-0.00$       & $-0.00$       & $0.01^{***}$  & $-0.01^{***}$ & $0.03^{***}$  \\
                             & $(0.00)$      & $(0.00)$      & $(0.00)$      & $(0.00)$      & $(0.00)$      & $(0.00)$      & $(0.00)$      & $(0.00)$      \\
HDD                          & $-0.00$       & $0.00$        & $-0.01^{***}$ & $0.00$        & $0.00$        & $-0.00^{***}$ & $0.01^{***}$  & $-0.02^{***}$ \\
                             & $(0.00)$      & $(0.00)$      & $(0.00)$      & $(0.00)$      & $(0.00)$      & $(0.00)$      & $(0.00)$      & $(0.00)$      \\
$\rho$                       & $0.01^{*}$    & $0.01^{**}$   & $0.01$        & $-0.00$       & $0.01^{**}$   & $0.02^{***}$  & $-0.00$       & $0.00$        \\
                             & $(0.00)$      & $(0.00)$      & $(0.00)$      & $(0.00)$      & $(0.00)$      & $(0.00)$      & $(0.00)$      & $(0.00)$      \\
\hline
Num. obs.                    & $222$         & $86$          & $50$          & $86$          & $222$         & $91$          & $84$          & $47$          \\
Parameters                   & $12$          & $12$          & $12$          & $12$          & $12$          & $12$          & $12$          & $12$          \\
Log Likelihood               & $-814.97$     & $-261.35$     & $-124.85$     & $-241.42$     & $-818.28$     & $-265.41$     & $-252.95$     & $-121.98$     \\
AIC (Linear model)           & $1656.09$     & $553.58$      & $273.78$      & $504.92$      & $1665.20$     & $575.09$      & $528.86$      & $266.13$      \\
AIC (Spatial model)          & $1653.95$     & $546.69$      & $273.71$      & $506.84$      & $1660.56$     & $554.82$      & $529.90$      & $267.96$      \\
LR test: statistic           & $4.15$        & $8.89$        & $2.07$        & $0.07$        & $6.64$        & $22.27$       & $0.96$        & $0.17$        \\
LR test: p-value             & $0.04$        & $0.00$        & $0.15$        & $0.79$        & $0.01$        & $0.00$        & $0.33$        & $0.68$        \\
\hline
\multicolumn{8}{l}{Statistical significance: \scriptsize{$^{***}pv<0.001$; $^{**}pv<0.01$; $^{*}pv<0.05$}}
\end{tabular}
\label{table:coefficients}
\end{center}
\end{sidewaystable}

\section{Conclusion}\label{sec6}

In this paper, we addressed the limitations of traditional regression models in handling spatial data, particularly when the relationships between variables exhibit heterogeneity across different clusters. Our motivation is based on the need to account for spatial autocorrelation in regression models with spatial clustering. Indeed, existing spatially clustered regression models do not consider spatial autocorrelation while estimating the parameters, albeit the spatial proximity of the units is enhanced by the algorithm.

To address this gap, we proposed the Spatially-Clustered Spatial Autoregression (SCSAR) model. This model extends the classical spatial autoregressive model by allowing the regression coefficients to vary spatially according to a cluster-wise structure. We also discussed extensions to other spatial models, such as the Spatial Error Model (SEM) and the Spatial Lagged Covariates (SLX) model. Our approach involves jointly estimating cluster memberships and regression coefficients through a penalized maximum likelihood algorithm, which encourages neighboring units to belong to the same spatial cluster with shared regression coefficients.

We applied the proposed SCSAR model to investigate the dynamic of statistical concentration of the European agricultural market concentration over the last decade. Using data on the number of active farms and their economic capacity (production/standard output) from 2010 and 2020, we assessed the local spatial spillovers of market concentration in European regions (NUTS-2 level). Our empirical results revealed heterogeneous local effects of explanatory variables on regional market concentration, identifying a clustering structure that partitions Europe into three groups: France and the Iberian Peninsula, Mittel-European and Scandinavian regions, and Eastern and Southern Europe regions. This application underscores the utility of our proposed methodology in uncovering hidden patterns and providing deeper insights into complex spatial data.

Future research could extend our work in several directions. Further methodological developments could focus on enhancing the robustness of the SCSAR model to various types of spatial weight matrices and exploring alternative penalty terms for spatial clustering. Additionally, another potential area for future research is the incorporation of dynamic structure to capture temporal changes in spatial clusters and regression relationships over time. Moreover, the development of methodologies to deal with panel data would be beneficial. Considering application-wise extensions, our model could be applied to other domains such as urban planning, environmental science, and public health to validate its effectiveness and uncover domain-specific insights. Finally, we highlight that investigating ways to address the curse of dimensionality and integrating LASSO-type methods for variable selection in the spatially-clustered context (especially when the number of covariates dramatically increases as in the caso of the SLX model) also represents a relevant topic for future studies.



\section*{Declarations}
\subsection*{Conflict of interest}
There is no conflict of interest to declare.
\subsection*{Data availability}
Data used in the paper are public (source Eurostat). Since we use only public data, no Special Permission is need to use copyrighted material from other sources (including the Internet). For reproducibility purposes, all scripts and the data are available at the following GitHub folder \href{https://github.com/PaoloMaranzano/RC_PM_RM_SCSAR_AgroConcentration.git}{https://github.com/PaoloMaranzano/RC\_PM\_RM\_SCSAR\_AgroConcentration.git}.


\bibliographystyle{unsrt}  
\bibliography{sn-bibliography}

\newpage

\appendix

\section{Appendix: other SCSAR models}
\label{appendix1}

\subsection{Spatially clustered regression with Spatial Error Model (SEM)}

\noindent An alternative model specification to account for spatial autocorrelation is the so-called spatial error model (SEM). The SEM model, 
\begin{equation}
\label{eq:sem}
    y_i = \sum_{p=1}^{P} \theta_p x_{ip} + \varepsilon_i, \quad \varepsilon_i = \lambda \sum_{j=1}^{N} w_{ij} \varepsilon_j + \eta_i,
\end{equation}
assumes a SAR structure for the error term. The parameter $\lambda$ defines the spatial autoregressive parameter and it is estimated with maximum likelihood. The joint log-likelihood function in the absence of clustering is given by
\begin{equation}
\label{eq:semloglik}
\begin{aligned}
    f_i(y_i| \mathbf{x}_i, \mathbf{W}, \boldsymbol{\delta}) &=-\frac{N}{2} \ln(2 \pi \sigma^2) + \ln|\det(I_N - \lambda \mathbf{W})| +\\
    - &\frac{1}{2 \sigma^2} \sum_{i=1}^{N} \left[y_i - \sum_{p=1}^{P} \theta_p x_{ip} - \lambda \sum_{j=1}^{N} w_{ij} \left(y_j - \sum_{p=1}^{P} \theta_p x_{jp}\right)\right]^2,
\end{aligned}
\end{equation}
where we define $\boldsymbol{\delta}=[\boldsymbol{\theta},\lambda]'$. In the case of cluster structure, the \eqref{eq:sem} can be written as
\begin{equation}
\label{eq:sem2}
    y_{ik} = \sum_{p=1}^{P} \theta_p x_{ip} + \varepsilon_{ik}, \quad \varepsilon_{ik} = \lambda_k \sum_{j=1}^{N} w_{i_k j_k} \varepsilon_{jk} + \eta_{ik}, \forall k=1,\dots,K,
\end{equation}
and the penalized log-likelihood to maximize is given by
\begin{equation}
\label{eq:func3b}
    Q(\boldsymbol{\delta},k) = \sum_{i=1}^{N} \log f_{ik} \left(y_i \mid \mathbf{x}_i, \mathbf{W},  \boldsymbol{\delta}_{k}\right)+\phi \sum_{i<j} w_{ij} I(k_i = k_j),
\end{equation}
with $\log f_{ik}$ being the \eqref{eq:semloglik} for the $k$-th cluster. The algorithm is the same as the one discussed in Table 1, alternating cluster assignment and the parameter estimation steps until convergence.

\subsection{Spatially clustered regression with Spatial Lag Model on covariates (SLX)}
The Spatial Lag of X (SLX) model extends the basic linear regression model by incorporating the spatial lags of the independent variables. The model can be expressed as
\begin{equation}
\label{eq:slx}
y_i = \sum_{p=1}^{P} \beta_p x_{ip}  + \sum_{j=1}^{N} w_{ij} \sum_{p=1}^{P}  \theta_p x_{jp} + \epsilon_i,
\end{equation}
where $y_i$ is the dependent variable for the $i$-th observation, $x_{ip}$ is the value of the $p$-th exogenous variable for the $i$-th observation, $\beta_p$ is the coefficient for the $p$-th exogenous variable, $w_{ij}$ is the $(i,j)$-th element of the spatial weight matrix $\mathbf{W}$, $x_{jp}$ is the value of the $p$-th exogenous variable for the $j$-th observation, $\theta_p$ is the coefficient for the spatially lagged $p$-th exogenous variable, and $\epsilon_i$ is the error term for the $i$-th observation, assumed to be iid $\epsilon_i \sim \mathcal{N}(0, \sigma^2)$.

The log-likelihood function for the SLX model, assuming normally distributed errors, is given by
\begin{equation}
\label{eq:slxloglik}
f_i(y_i| \mathbf{x}_i, \mathbf{W}, \boldsymbol{\delta}) = -\frac{N}{2} \log (2 \pi \sigma^2) - \frac{1}{2 \sigma^2} \sum_{i=1}^{N} \left( y_i - \sum_{p=1}^{P} x_{ip} \beta_p - \sum_{j=1}^{N} w_{ij} \sum_{p=1}^{P} x_{jp} \theta_p \right)^2,
\end{equation}
where $\boldsymbol{\delta}=[\boldsymbol{\beta},\boldsymbol{\theta}]'$.Therefore, the SLX model can be seen as a standard regression model with lagged covariates. The solution of the maximum likelihood estimation is very similar to the standard regression model and, therefore, the SLX model can be easily estimated using the \cite{sugasawa2021spatially} spatially clustered algorithm. The clustered version of \eqref{eq:slx} can be written as

\begin{equation}
\label{eq:slxb}
y_{ik} = \sum_{p=1}^{P} \beta_{pk}  x_{ipk} + \sum_{j=1}^{N} w_{ij} \sum_{p=1}^{P} \theta_{pk} x_{jpk}  + \epsilon_{ik}, \quad \forall k=1,\dots,K
\end{equation}
The objective function to maximize to obtain a spatially clustered regression model is
\begin{equation}
\label{eq:func4}
    Q(\boldsymbol{\delta},k) = \sum_{i=1}^{N} \log f_{ik} \left(y_i \mid \mathbf{x}_i, \mathbf{W},  \boldsymbol{\delta}_{k}\right)+\phi \sum_{i<j} w_{ij} I(k_i = k_j),
\end{equation}
with $\log f_{ik} \left(y_i \mid \mathbf{x}_i, \mathbf{W},  \boldsymbol{\delta}_{k}\right)$ being equal to \eqref{eq:slxloglik} for the $k$-th cluster. The algorithm is therefore the same as Table 1, alternating parameter estimation and cluster assignment until convergence.

\textbf{Remark}. We remark that the SLX model is more subject to the course of dimensionality. Indeed, while in the standard regression model, we deal with $P$ exogenous variables, due to the spatial lags the variables in the SLX model increase quadratically. While spatially clustered models are considered, the parameters are estimated within each cluster using $N_k < N$ clustered observations. This means that the ratio $N_k/P$ becomes much lower and the parameter estimation step becomes unreliable with not large enough $N_k$ observations. Therefore, we suggest that future research should focus on the development of LASSO-type (e.g. see \cite{di2023lasso}) spatially clustered SLX methods to cope with the course of dimensionality in this setting.

\end{document}